# Predictive Real-Time Control Optimization of a Stormwater Management System*

Shishegar, S.[1], Duchesne, S.[2], and Pelletier, G[3].

*Abstract*— An optimization model of a stormwater pond is developed to improve the performance of the system in terms of water quantity and quality. Nowadays, stormwater management systems play an important role in mitigating the impacts of urbanization on the natural hydrological cycle. These systems can be managed in such a way that they meet smart city needs. An automated dynamically managed system that can adapt itself to ever-changing environmental conditions can be modeled using a mathematical optimization approach. Hence, a Predictive Real Time Control (PRTC) approach is proposed in this paper to optimize the performance of stormwater management basins in terms of minimizing hydraulic shocks during wet periods. Then some generalized rules are designed to control the sedimentation of trapped water in the pond during dry periods to improve the quality of water discharged to the receiving stream. The combination of these two approaches provides a real-time improved performance in comparison to the static strategy. *Keywords:* Optimization model, Real time control, Stormwater management, Hydraulic shocks, Water quality

I. INTRODUCTION

The characteristics of rainfall trends are changing in different parts of the world. It has been shown that the annual precipitation in Canada has increased over the past 50 years. However, modifications in intensity of extreme rainfall events are recognized as the most significant consequence of Climate Change [1]. This issue along with increased urbanization can raise environmental and public health concerns in urban areas. Hydraulic shocks (i.e. sudden elevations in water pressure due to the velocity sudden changes) to the receiving river and degradation of water quality caused by stormwater runoff that reaches the outlet of the sewer network are two issues that should be controlled. Existing infrastructure placed in most urban areas are not equipped for these controls and they need an upgrade to adapt to the changing environmental conditions. Stormwater basins (or ponds) are one of the stormwater management solutions that has been developed to control flow rates and water quality. However, they are generally designed and operated statically based on the analysis of past rainfall data. While, to design a sustainable and adaptive system to the changing environmental conditions like Climate Change and urbanization, predictive control strategies, which consider future weather conditions based on new technologies should be taken into account. In fact, advances in technology and automatic systems can lead to a new generation of "automated" stormwater systems, which perform predictive control to instantly modify themselves to adapt to changing inputs [2]. This could be realized by employing Real Time Control (RTC) strategies for city's infrastructures. These strategies can be defined as controls that monitor process variables at the system and generate set points for the actuators [3] using the data provided by the sensors installed over the system and implemented by the controllers that are responsible of translating the output set points into actions. Now if this dynamically managed system considers predicted data, besides actual and historical data, it could become capable of adapting itself to instability in its environment's conditions. Optimization methods are strong tools to realize Predictive Real Time Control (PRTC) of existing stormwater basins [4] for an adaptive management of urban stormwater. Therefore an optimization model is developed in this study to provide an optimal performance for the stormwater basins by considering outflow as a decision variable. This optimization would be executed during wet periods, followed by an operation based on some generalized rules during dry periods to control water sedimentation in the basins. This could help water discharge to the receiving watercourse with a better quality.

The objectives of this paper are to:
• Evaluate the performance of predictive real time control optimization strategies of stormwater management basins in terms of minimization of the hydraulic shock imposed on the river during wet periods.
• Demonstrate the sedimentation performance of stormwater basins managed by some designed quality control rules during dry periods, and
• Study the impact of using the combination of PRTC strategy and quality control rules on preventing overflows of stormwater basins.

The structure of the paper is as follows; section II reviews briefly the literature related to the automated stormwater systems. In section III, we present the methodology to develop a combined PRTC optimization model and quality control rules, and then the results are discussed in section V. Finally, the last section provides the conclusion of this study.

*Research supported by Fonds de Recherche du Quebec- Nature et Technologie (FRQNT).

1-Shadab Shishegar is PhD candidate at Institut National de la Recherche Scientifique, 490 rue de la Couronne (phone: +1(418)654-4677; e-mail: shadab.shishegar@ ete.inrs.ca).
2-Sophie Duchesne is professor at Water, Earth and Environment Center of Institut National de la Recherche Scientifique, 490, rue de la Couronne Québec (Québec), G1K 9A9 Canada (e-mail: sophie.duchesne@ete.inrs.ca).
3-Geneviève Pelletier is professor at Department of Civil and Water Engineering, Faculty of Science and Engineering, Université Laval, Pavillon Adrien-Pouliot 1065, av. de la Médecine, Québec (Québec), G1V 0A6, Canada (e-mail: Genevieve.Pelletier@gci.ulaval.ca).

## II. LITERATURE REVIEW

RTC of stormwater basins can lead to a sustainable and adaptive management of urban stormwater. Unlike statically controlled stormwater management infrastructure, which cannot adapt their operation to different storm events or changing land uses, RTC systems can use integrated coordination to control water quality and hydraulic impacts at stormwater management sites, in particularly ponds and basins [5]. Few studies employed optimization methods to achieve the optimal performance of stormwater management systems in real time. In this regard, a generalized reduced gradient optimization algorithm was proposed by Duchesne et al. in 2004 to minimize overflow volumes from combined sewers during wet periods [6]. The global and predictive RTC strategy presented in the study determines the intercepted flows in such a way that surcharge is allowed to achieve the optimal use of storage and transport capacity of the sewer system. The rolling horizon approach was employed for RTC. On the other hand, in related literature, there are studies that proposed rule-based strategies which have shown improvement of the system performance like the one presented in [7]: by designing an adjustable gate opening, a flexible structure is provided in order to adjust the outflow rate of a single stormwater basin to increase the Total Suspended Solids (TSS) removal efficiency while decreasing the hydraulic shock to the receiving water bodies. To do so, some enhanced RTC rules were designed that finally improved the pollutant removal performance efficiency of the basin to about 44%. However, the rules were defined by trial-and-error for this specific problem, so they cannot simply be applied on different types and sizes of basins. Having generalized rules that are applicable to all stormwater basins submitted to various types of weather conditions would be more advantageous. Furthermore, considering that the outlet control of the stormwater ponds could be a key factor to improve the system performance, integrating optimization methods in the control could even provide an optimal performance for the stormwater management systems in terms of minimizing hydraulic shocks on the receiving water bodies [8].

Recently, benefits achieved by the implementation of a combination of local rule-based controls with some higher-level optimization algorithms, in order to manage the urban drainage systems, have been reported [9]. While these strategies have shown improved performance for the control of large sewer networks, developing some efficient RTC optimization strategies for the application of stormwater basins has not yet been achieved.

In the present work, we develop a mathematical optimization model for dynamic control of stormwater discharges from a stormwater basin to provide a real time outflow scheduling based on the present and future weather conditions. Further, we explore the performance of this stormwater basin in terms of pollutant removal efficiency through sedimentation by designing some control rules in order to allow the pond to properly empty for the next storm event during dry periods. The combination of these two approaches is expected to improve the performance of the stormwater pond, as opposed to a static control, in terms of quality and quantity. The contribution of this study is the integration of logic-based rules in the optimization model where the optimization part minimizes the hydraulic shocks to the river by the determination of optimal outflow set points for the basin, then the rule-base part aims at controlling the sedimentation of the retained water while looking at the required volume for the upcoming inflow to the basin. The combination of these two strategies will finally enhance the ability of the basin to meet stormwater management regulations, provided that the basin is retrofitted by some commonly used equipment like sensors, control valves and a moveable gate to apply the aforementioned methodology. This would transform an infrastructure that had been managed traditionally so far to a smart infrastructure.

## III. METHODOLOGY

First, an optimization algorithm for the PRTC of the stormwater basins is developed. The methodology is started by modelling the problem mathematically by defining the objective function, which minimizes hydraulic shocks in the receiving water course. Given the predictive nature of the RTC strategies, decisions made for manipulating the outlet gates will take into account not only the current weather conditions but the future conditions, based on short-term weather forecasts. Hence, the required data includes: i) observed and predicted rainfall data; ii) characteristics of the studied watershed and its basins like slopes, area, imperviousness, etc.; iii) hydrological and hydraulic models to transform these data into water flows, runoff volumes, water levels either that of the river or the basins, etc.; and iv) an optimization algorithm to define the management rules based on the predicted and observed state of the system. Using these data, the optimization of the designed objective function is done subject to the problem constraints, both of which can be simulated through the Stormwater Management Model-SWMM [10]. SWMM dynamically simulates stormwater runoff and flow in stormwater sewer networks from specified rainfall series. This process provides the inflow hydrographs of the studied basin, which is used as input for the optimization model. While optimizing the system in wet periods, outflows at the outlet of the stormwater basin is determined to minimize the outflow to the river. Then, during the dry periods, three designed rules provide the outflows in such a way that sedimentation is maximized and, consequently, water is discharged into the river with a better quality. This process is continued as long as the termination condition is not satisfied.

### A. Optimization Model for Quantity Control

The PRTC optimization model is developed as follows:

$$Min \sum_{t=0}^{n_c} \{Q(t)\} \quad (1)$$

Where:
$Q(t)$ = The outflow decision variable from the stormwater basin at time step $t$ (m$^3$/s).
$n_c$ = Number of time steps of the control horizon.

This function aims at minimizing the total outflows from the basin at time step $t$, which minimizes the hydraulic shocks imposed on the receiving river, as one of the important functions of urban stormwater basins. The constraints of our PRTC optimization model are:

- *Capacity constraint*

$$\sum_{t=0}^{n_c}(I(t) - Q(t))\Delta t \le A * H_{max} \quad (2)$$

Where:
$I(t)$ = The inflow to the basin at time step $t$ (m³/s).
$A$ = Area of the studied basin (m²).
$H_{max}$ = Height of the basin (m).
$\Delta t$ = Difference of $t$ between two time steps (s)

- *Mass balance constraint*

$$\begin{aligned}Q(t)\Delta t + 2A * H(t) \\ = I(t-1)\Delta t + I(t)\Delta t + 2A * H(t-1) \\ - Q(t-1)\Delta t \\ \forall t = 1, \dots, n_c\end{aligned} \quad (3)$$

Where:
$H(t)$ = The water depth dependent decision variable in the basin at time step $t$ (m).

It should be noted that Equation 3 results from the hydraulic routing equation $\Delta V/\Delta t = \bar{I} - \bar{Q}$.

Also as decision variable $H(t)$ is in conflict with $Q(t)$ in nature (having more outflow leads to have less water in the basin), this constraint balances the water volume and flow in the basin. Actually, considering the limited capacity of the stormwater pond and also its function to release the trapped water slowly to the receiving watercourse, the presence of this mass balance constraint in the optimization model is inevitable.

- *Positivity constraint*

$$H(t) + Q(t) > 0 \quad \forall t = 0,1,\dots,n_c \quad (4)$$

This constraint confirms that the optimization model is in process only when there is an inflow to the basin i.e. during the wet period.

- *Outflow constraint*

$$0 \le Q(t) \le Q_{max} \quad \forall t = 0,1,\dots,n_c \quad (5)$$

Where:
$Q_{max}$ = The maximum allowable outflow from the basin (m³/s).

- *Non-negativity constraint*

$$H(t) \ge 0 \quad \forall t = 0,1,\dots,n_c \quad (6)$$

B. *Designed Rules for Quality Control*

The existing strategies in the literature for controlling the sedimentation in stormwater basins resulted from trial-and-error to come up with a threshold to differentiate between rules like the one in [7]. However, the strategy that we propose here are generalized rules that could be used in different cases with different climate conditions and rainfall series. These designed rules consider the next rainfall event along with the required storage volume in the basin to treat the upcoming inflows resulting from this rainfall event. Since the desired minimum retention time is 20 h to decrease significantly the TSS concentration in water according to [11], this time is combined with the emptying time of the basin to specify the emptying rule. Then the selected rule provides the outflow from the basin as a result of manipulating the opening of the outflow gate. Equations 7, 8 and 9 present the proposed rules.

$$\begin{aligned}&if\ t_{next\ rain} \le t_e \rightarrow \\ &Q(t) = Q_{max}\end{aligned} \quad (7)$$

$$\begin{aligned}&if\ t_e < t_{next\ rain} \le t_e + 20h \rightarrow \\ &Q(t) = Q_{max} * \frac{t_e - t_f}{t_{next\ rain} - t_f}\end{aligned} \quad (8)$$

$$\begin{aligned}&if\ t_e + 20h < t_{next\ rain} \rightarrow \\ &Q(t) = 0\end{aligned} \quad (9)$$

Where:

$t_e$ = Emptying time of the basin until availability of the required storage volume (s) and

$$t_e = H(t) * A/H_{max} * 360 \quad (10)$$

$t_{next\ rain}$ = Remaining time to the start of the next predicted rainfall event (s).

$t_f$ = The time that the previous rainfall event is finished (s).

C. *Rolling horizon*

To realize the RTC and execute the optimization-simulation periodically, the rolling horizon approach is employed. This approach is based on the planning for a few time periods ahead, and then expanding the time horizon at each step, after receiving feedbacks from the system following the implementation of the previously determined set points. This concept is shown schematically in Fig. 1. This approach simulates the problem in a way that when the real data become available, they can be put into the model to update the information and conclusively re-plan for the next remaining time periods. This is advantageous for problem solvers, to solve a large problem in multiple phases, instead of solving it thoroughly, which may become too time-intensive. An example of implementing this rolling horizon approach, which inspired the approach used in the presented study, can be found in [6].

The planning horizon in our case is 60 hours, in which the periods are divided in $n_c$ optimization time steps. The simulation is performed in 5-minute time steps. It means that the data are updated each 5 minutes based on the observations and their comparison with the predicted data.

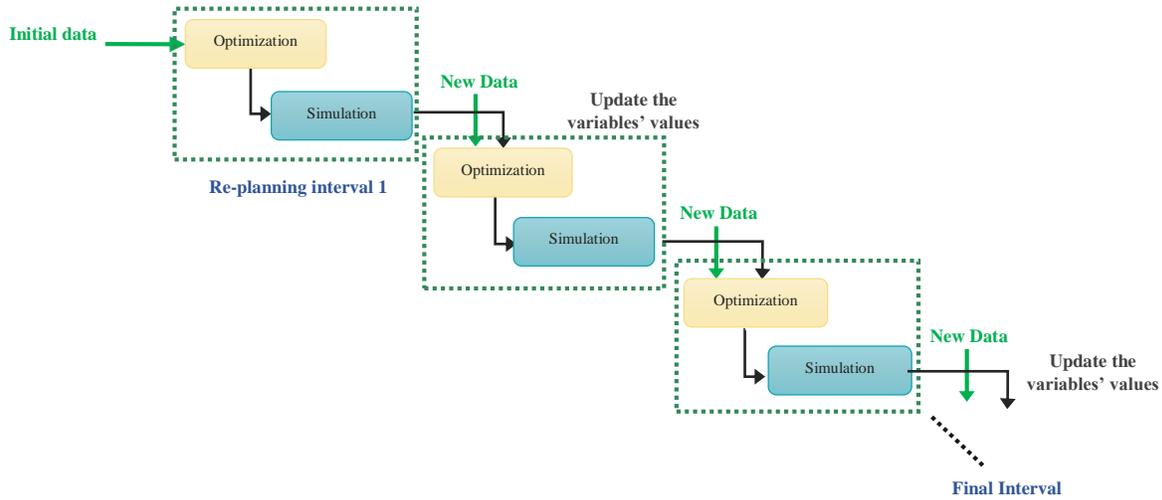

Figure 1-Simulation and optimization interaction in rolling horizon approach

The planning in $n_c$ periods is rolled over the total planning horizon (so called anticipation horizon), until it reaches the final period.

Based on this methodology, three steps are followed for each rainfall events:

1. The receiving stormwater runoff is trapped in the basin as long as the PRTC decides in order to intenerate sharp discharges into the river and also contain possible first-flush flows.

2. The outflow rates are determined by the optimization algorithm in such a way that the hydraulic shock on the receiving water is minimized during wet period while preventing the basin to overflow.

3. Finally, the retention time of the basin is defined by the proposed rules so that the quality of discharged water is improved after the rainfall event.

## IV. RESULTS AND DISCUSSION

A stormwater pond located at the outlet of a stormwater sewer network of a city situated in the southern region of the province of Quebec (Canada) was chosen as the case study to validate the proposed strategy. This sector is known for its high possibility to be flooded due to stormwater runoff and possible hydraulic shocks induced to the receiving watercourse. Furthermore, the stormwater runoff in this area has a relatively high concentration of sediments which getting even more polluted during wash off of impervious surfaces.

Testing the proposed strategy on this stormwater pond resulted in the following hydrographs (Fig. 2). The inflow shown in Fig. 2 has been obtained using a 60-hour period extracted from a one-year time-series rainfall data of a close region in the year 2013. As the prediction data was not available, the exact prediction is considered in our case in such a way that the real-time reception of the future data in 5-minute time steps is assumed. Table 1 presents the values of the parameters of the case study basin. As shown in hydrograph of Fig. 2, the optimization-simulation is executed periodically to realize the RTC control during the wet periods, i.e. when there is an inflow to the basin, the basin retains the water as much as the optimization model decides, while during dry periods, the basin gets prepared for the next rainfall event via the quality control rules designed in this regard. If the next rainfall is close, the outlet gate is manipulated so that the required storage volume for the next coming rainfall becomes available, else the trapped water discharges into the river when its TSS pollutants are settled enough. On the other hand, as illustrated in Fig. 2, when the outlet gate is closed and there is no close predicted rainfall, the trapped water volume in the basin increases. This is where the PRTC model decides on a flow rate that avoids any overflows and releases the water slowly. For example, at time step 61, a sharp inflow enters the basin; under RTC control, the optimization model decides to open the gate to allow the maximum outflow while in static control, a sharp outflow of 13.2 $m^3$/s from the basin discharges to the river. As mentioned before, these high discharges are the main cause of hydraulic shocks in the receiving streams. Also, during the dry period coming right after the rain event started at time-step 61, the quality control rules decide to keep the remaining water by closing the outlet gate for 20 h to allow sedimentation, through which the TSS load is reduced [11] and the water quality is improved in order to be discharged to the receiving watercourse. Then at time step 305, after 20 h of retention, the gate is opened slightly to let the retained water released to the river. This decision is taken as there is no rainfall event predicted in the near future. While in time step 605, although the water is retained only during 11 h, since the upcoming rain is predicted for less than $t_e$, an opening of the outlet allows the required volume to be emptied from the basin to catch the incoming inflows.

TABLE I.  CASE STUDY STORMWATER BASIN PARAMETERS VALUE

| Parameter | | | |
|---|---|---|---|
| Volume ($m^3$) | Max. height (m) | Max. outflow ($m^3$/s) | Num. of time steps |
| 61495 | 1.2 | 2.54 | 720 |

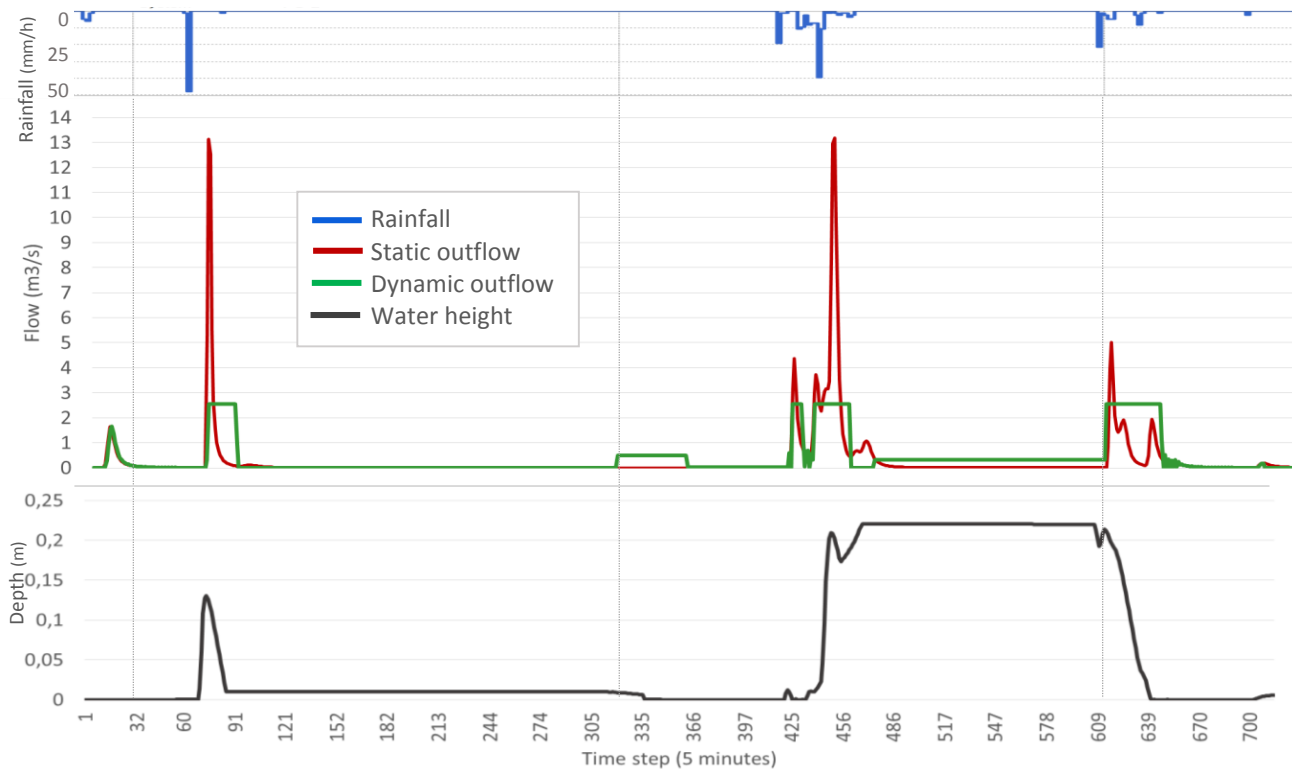

Figure 2-Hydrographs related to the basin under static vs. dynamic controls during dry and wet periods

As a result, it is observed that the studied stormwater basin performs in such a way that it achieves all its operational objectives, i.e. employing the combined PRTC and rule-based strategies helped to get scheduled outflows that impose the least hydraulic shock to the receiving stream while preventing overflow of the basin. On the other hand, the controlled retention time of water provides enough time for sedimentation and improves the pollutant removal efficiency of the stormwater management basin.

## V. Conclusion

This study shows how predictive real time control optimization of a stormwater basin combined with generalized rule-based strategies can satisfy stormwater management goals like quantity control (via minimizing hydraulic shocks to the receiving watercourse) and quality control (by improved sedimentation). This approach provides an efficient tool that generates flow rates at the outlet of the basin in such a way that while emptying the basin, it avoids inducing sharp discharges to the downstream watercourse. Similarly, this active regulation of outflow rates, also improves water quality by enhancing the removal of its contaminants and mitigating the non-point source pollutions carried by stormwater runoff.

In conclusion, employing this strategy alongside with retrofitting the statically controlled stormwater management systems with some commonly available equipment, we can exploit the system's complete potential and achieve an automated infrastructure with an optimal quality and quantity control performance, which is sustainable and adaptive to the changing environment conditions.